\input harvmac
\input epsf

\newcount\figno
\figno=0
\def\fig#1#2#3{
\par\begingroup\parindent=0pt\leftskip=1cm\rightskip=1cm\parindent=0pt
\baselineskip=11pt
\global\advance\figno by 1
\midinsert
\epsfxsize=#3
\centerline{\epsfbox{#2}}
\vskip 12pt
{\bf Fig. \the\figno:} #1\par
\endinsert\endgroup\par
}
\def\figlabel#1{\xdef#1{\the\figno}}
\def\encadremath#1{\vbox{\hrule\hbox{\vrule\kern8pt\vbox{\kern8pt
\hbox{$\displaystyle #1$}\kern8pt}
\kern8pt\vrule}\hrule}}

\overfullrule=0pt

\noblackbox
\parskip=1.5mm

  
\def\npb#1#2#3{{\it Nucl. Phys.} {\bf B#1} (#2) #3 }
\def\plb#1#2#3{{\it Phys. Lett.} {\bf B#1} (#2) #3 }
\def\prd#1#2#3{{\it Phys. Rev. } {\bf D#1} (#2) #3 }
\def\prl#1#2#3{{\it Phys. Rev. Lett.} {\bf #1} (#2) #3 }
\def\mpla#1#2#3{{\it Mod. Phys. Lett.} {\bf A#1} (#2) #3 }
\def\ijmpa#1#2#3{{\it Int. J. Mod. Phys.} {\bf A#1} (#2) #3 }

\def\cmp#1#2#3{{\it Commun. Math. Phys.} {\bf #1} (#2) #3 }

\def\prep#1#2#3{{\it Phys. Rep.} {\bf #1} (#2) #3 }


\def\half{{1\over 2}\,}


 \def\CH{{\cal H}}


\lref\rostr{A. Sagnotti, \prep{184}{1989}{167\semi}
P. Horava, \plb{231}{1989}{251;} \npb{327}{1989}{461\semi}
M. Bianchi and A. Sagnotti, \plb{247}{1990}{517\semi}
M. Bianchi, G. Pradisi and A. Sagnotti, \npb{376}{1992}{365.}
}
\lref\rgreen{M.B. Green, \npb{381}{1992}{201.}}
\lref\ratickw{J. Atick and E. Witten, \npb{310}{1988}{291.}}
\lref\ralos{E. \'Alvarez and M.A.R. Osorio, \npb{304}{1988}{327.}}
\lref\rbranva{R.H. Brandenberger and C. Vafa, \npb{316}{1988}{391.}}
\lref\rpol{J. Polchinski, \cmp {104}{1986}{37.}} 
\lref\rgr{M.B. Green, \plb{282}{1992}{380.}}
\lref\rb{J.L.F. Barb\'on, Preprint PUPT-96-1635, to appear.}
\lref\rrev{J. Polchinski, S. Chaudhuri and C. Johnson, Preprint NSF-ITP-96-003,
hept-th/9602052.}
\lref\rao{E. \'Alvarez and M.A.R. Osorio, \prd{36}{1987}{1175.}}
\lref\rpcmp{J. Polchinski, \cmp{104}{1986}{37.}}
\lref\rtdual{
K.H. O'Brien and C.I. Tan, \prd {36}{1987}{1184\semi}
B. McClain and B.D.B Roth, \cmp{111}{1987}{539\semi}
E. \'Alvarez and M.A.R. Osorio, \prd{40}{1989}{1150.} }
\lref\rrohm{R. Rohm, \npb{237}{1984}{553.}}
\lref\ros{M.A.R. Osorio, \ijmpa{7}{1992}{4275.}}
\lref\rsdual{E. Witten, \npb{443}{1995}{85.}}
\lref\rpolc{J. Polchinski, \prl{75}{1995}{4724.}}
\lref\rdir{M.B. Green, \plb{266}{1991}{325.}}
\lref\rseipol{J. Dai, R.G. Leigh and J. Polchinski, \mpla
{4}{1989}{2073\semi}
R.G. Leigh, \mpla{4}{1989}{2767.}}
\lref\rkap{J.I. Kapusta, {\it Finite-Temperature Field Theory}, Cambrige 1989.}
\lref\rcaipol{J. Polchinski and Y. Cai, \npb{296}{1988}{91.}}
\lref\rwguo{A. Erdelyi, {\it Higher Transcendental Functions}, McGraw-Hill,
New York 1953.}
\lref\rdoljack{L. Dolan and R. Jackiw, \prd{9}{1974}{3320.}}
\lref\raw{J. Atick and E. Witten, \npb{310}{1988}{291.}}


\line{\hfill IASSNS-HEP-96-73}
\line{\hfill hep-th/9607052}
\vskip 0.5cm

\Title{\vbox{\baselineskip 12pt\hbox{}   }}
{\vbox{\centerline{Open String Thermodynamics and D-Branes}
\bigskip
}}
\centerline{M. A. V\'azquez-Mozo\foot{E-mail: vazquez@sns.ias.edu}}
\smallskip
\centerline{\it School of Natural Sciences}
\centerline{\it Institute for Advanced Study}
\centerline{\it Princeton, NJ 08540, USA}
\vskip 0.3in

We study the thermodynamics of open superstrings in the presence of
$p$-dimensional D-branes. We get some finite temperature dualities
relating the one-loop canonical free energy of open strings to the 
self-energy of D-branes at dual temperature. For the open bosonic string 
the inverse dual temperature is, as expected, the dual length 
under T-duality, $4\pi^{2}\alpha^{'}/\beta$. On the contrary, for the 
$SO(N)$, type-I superstring the dual temperature is given by $\beta$-duality, 
$2\pi^{2}\alpha^{'}/\beta$. We also study the emergence of the
Hagedorn singularity in the dual description as triggered by the 
coupling of the D-brane to unphysical tachyons as well as the high
temperature limit.


\Date{July, 1996}


\newsec{Introduction}

Apart from a few exceptions \refs\rostr,
open superstring theory has been left in a state of semi-oblivion 
during the last ten years mainly due to the good phenomelogical prospects of 
closed superstring models. However, after the avalanche of results following
the {\it second superstring revolution} in 1995 it has become clear
that open strings play an important r\^ole in the non-perturbative dynamics
of closed string models. At this moment there are solid evidences that
the weakly coupled $SO(32)$, type-I superstring describes the strong 
coupling regime of the $SO(32)$ heterotic string \refs\rsdual. Dirichlet 
open string theory \refs\rdir\ has also been incorporated into this 
landscape after the result of Polchinski \refs\rpolc\ that D-branes 
\refs\rrev\ carry R-R charge and therefore are candidates for the solitonic 
states demanded by S-duality of the type-IIB superstring \refs\rsdual. 

Open superstrings at finite temperature is an appealing scenario 
which should be re-explored in the light of the new results 
\refs\rgr\rgreen\rb . One interesting issue to study is the physics of 
strings at the Hagedorn temperature. Since the Hagedorn divergence
indicates a breakdown of the perturbative formalism at high temperatures
it is possible that the brand new string dualities could be of any use in 
deciding whether this signals any kind of phase transition or 
a maximum temperature of the string ensemble. The main obstacle in addressing
this problem is the breaking of supersymmetry at non-zero temperature;
consequently the application of zero temperature dualities 
in this setup is highly delicate. 

In the present note we are going to be concerned with
another interesting problem in the subject of thermal strings: 
finite temperature duality \refs\rtdual\raw.
It is well known that the partition function of the ten-dimensional heterotic 
string enjoys a $\beta$-duality\foot{In this paper I will use the
term $\beta$-duality to label the transformation $R\rightarrow \alpha^{'}/2R$
in opposition to T-duality, $R\rightarrow \alpha^{'}/R$, with $R$ the 
compactification radius.} relation connecting low and high temperatures 
($\beta=T^{-1}$)
$$
Z_{\rm het}(\beta)=Z_{\rm het}\left({2\pi^{2}\alpha^{'}\over \beta}\right).
$$
It is important to stress that the $\beta$-duality of the heterotic string 
differ from the usual T-duality by a factor two.
A consequence of this formal relation is the existence of a well-defined
high-temperature phase (beyond the Hagedorn point) with {\it anomalous}
thermodynamical properties \refs\ralos\raw. There are, however, no such 
self-duality for the type-II \refs\ros\ and type-I superstrings.

In type-I models strings and D-branes play a dual r\^ole under 
T-duality \refs\rseipol; therefore it is likely for a dual temperature 
description of a gas of open strings to involve D-branes. Of course this 
has to be true for the bosonic string, since in this case
finite temperature duality is just T-duality.
For open fermionic strings, on the other hand, the boundary conditions of
the space-time fermions makes the situation more involved and one must
proceed with some care. In the next section we will study
the canonical free energy of open superstrings and D-branes and, going
to the closed string channel, will obtain some duality relations between
the gas of strings and a number of ``hot'' semiclassical D-branes. To make
the exposition lighter, we will present in some detail the case of the 
bosonic string to center ourselves later in the much more interesting case of 
the $SO(N)$ open superstring. In section 3 we will look at the Hagedorn 
transition from the D-brane perspective and the high temperature limit of 
open strings will be study. Finally, in section 4 we will sumarize our 
conclusions.

\newsec{Open strings and D-branes at finite temperature}

Our first task will be to study a thermal ensemble of open strings in 
the presence of a $p$-dimensional D-brane. The best way to get the one-loop 
canonical free energy of such a system is to begin with a gas of open strings 
in a $D-1$ toroidal space\foot{$D=10$ for the superstring and $D=26$ for 
the bosonic string.} with radii $R_{i}$ ($i=1,\ldots,D-1$) and then take 
$R_{1},\ldots,R_{p}\rightarrow \infty$ at the same time than $R_{p+1},
\ldots,R_{D-1} \rightarrow 0$. The euclidean time is compactified at a 
fixed length $\beta=2\pi R_{0}=T^{-1}$ and target fermions are taken to be
antiperiodic in this circle. The one-loop canonical free
energy can be easily computed by summing over the individual contributions
of all the fields in the string taking into account their boundary conditions
\refs\rpcmp\rao\
$$
\log Z(\beta)= -{\rm Tr}_{\CH}\left[(-1)^{F}\int_{0}^{\infty} {dt\over 2t}
e^{-2\pi \alpha^{'} t(k^{2}+M^{2})}\right]+{\rm counterterms},
$$
where the trace is over the open string Hilbert space and $k^{2}$, $M^{2}$ are
respectively the momentum and mass operators and $F$ is the target space
fermion number. The counterterms remove the
vacuum energy in the zero-temperature limit $\beta\rightarrow \infty$ whenever
it is non-vanishing. Proceeding as described above we find for the bosonic
string without Chan-Paton factors\foot{In what follows we will be rather
cavalier in dealing with the standard divergences in the bosonic string 
amplitudes since they are a mere artifact of the tachyonic ground state. In any
case these kind of divergences are irrelevant for most of our analysis.}
\eqn\bos{
\log Z_{p}(\beta)=-\beta \int_{0}^{\infty} 
{dt\over 2t} 
(8\pi^{2}\alpha^{'}t)
^{-{p+1\over 2}}[\eta(it)]^{-24}\left[\theta_{3}\left(0\left|{i\beta^{2}\over 
8\pi^{2}\alpha^{'}t}\right.\right)-1\right].
}
In the case of the
$SO(N)$, type-I superstring the result is
\eqn\fer{
\eqalign{\log Z_{p}(\beta)&=-
{N^{2}\over 2}\int_{0}^{\infty} {dt\over 2t} 
(8\pi^{2}\alpha^{'}t)
^{-{p\over 2}}{\theta^{4}_{2}(0|it)\over 2\eta^{12}(it)}
\theta_{4}\left(0\left|{2i\pi^{2} \alpha^{'}t\over 
\beta^{2}}\right.\right)  \cr
&+{N\over 2}\int_{0}^{\infty} {dt\over 2t} 
(8\pi^{2}\alpha^{'}t)^{-{p\over 2}}
{\theta^{4}_{2}(0|it+1/2)\over 2\eta^{12}(it+1/2)}
\theta_{4}\left(0\left|{2i\pi^{2} \alpha^{'}t\over 
\beta^{2}}\right.\right).}
}
The first term represents the contribution of the annulus and the second
one that of the M\"obius strip. In both cases $(1/\beta)\log Z_{p}(\beta)$  
can be interpreted as the canonical free energy of a gas of open strings 
living in the $(p+1)$-dimensional world-volume of the D-brane in which
their endpoints live. Nevertheless, since open strings can fluctuate in the 
directions transverse 
to the D-brane they probe the full $D$-dimensional space as it must be since 
we are dealing with a critical string. Actually, it is interesting to notice 
that the only
way to get a {\it purely} $(p+1)$-dimensional string theory would be
to take the D-brane tension $T_{p}$ to infinity while freezing the string
tension at the same time. However this is not possible, since 
$T_{p}$ scales as $T_{p}\sim (\alpha^{'})^{-p+(D-4)/2}$. Thus we cannot
take $T_{p}\rightarrow \infty$ without at the same time taking $\alpha^{'}$ 
either to zero or to infinity (in the marginal case $p=(D-4)/2$ the D-brane
tension is a fixed numerical constant).

Up to now we have used the open string point of view. It is 
interesting, however, to look at the same problem from the D-brane perspective.
What we have now, instead, is a D-brane located at a fixed point of 
the space which couples to closed strings. At zero temperature the 
vacuum energy can be viewed as due to the emision and absorbtion of
(bosonic) closed string states. For the bosonic string this amounts to 
a divergent result, while for superstrings, on the contrary, the 
total vacuum energy is zero due to the cancellation between the 
contributions of R-R and NS-NS states. This is viewed in the open 
string channel as the result of the vanishing of the one-loop 
cosmological constant for the type-I superstring.

At non-zero temperature supersymmetry is broken in the open string channel
of the supersymmetric string
and therefore there is no non-renormalization theorem which prevents us
from having a net vacuum energy. Going to the closed string channel this
means that the contributions to the self-energy of the D-brane coming from
R-R and NS-NS states do not cancel any more and thus they are 
weighted differently in the sum over closed string states. In what follows
we will construct the prescription to compute this ``thermal'' self-energy of
the D-brane. 

In order to illustrate the techniques to be used here we will work out first 
in some detail the case of the oriented bosonic string, in which we already 
know the kind of results to be expected. We can rewrite \bos\ in the
closed string channel by changing to a new integration variable $s=t^{-1}$
(for the time being we will include the divergent vacuum energy)
$$
Z_{p}(\beta)=-\half\beta (8\pi^{2}\alpha^{'})^{-{p+1\over 2}}\int_{0}^{\infty}
ds\, s^{-{25-p\over 2}}[\eta(is)]^{-24}\theta_{3}\left(0\left|{i\beta^{2} 
s\over
8\pi^{2}\alpha^{'}}\right.\right).
$$
Now we use the expansion of the Dedekind eta function
$$
[\eta(is)]^{-24}
=\sum_{N=0}^{\infty}c(N) e^{-2\pi s(N-1)} \hskip 1cm c(N)\in {\bf Z}
$$
and after trading variables again in the integral, $s\rightarrow 
s'=\pi\alpha^{'}s/2$, we have
$$
Z_{p}(\beta)=-{1\over 8} (4\pi)^{p-25\over 2}
\left[{\sqrt{\pi}\over 16}(4\pi^{2}\alpha^{'})^{
11-p\over 2}\right]^{2}\sum_{N=0}^{\infty}c(N)\sum_{m\in{\bf Z}}
\int_{0}^{\infty} ds\, s^{-{25-p\over 2}} e^{-s\left[4\pi^{2}m^{2}\beta^{2}+
{4\over \alpha^{'}}(N-1)\right]};
$$
the term between square brackets gives the $p$-brane tension;
in the exponent it is easy to recognize the mass formula for the closed
string with $N$ the oscillator level of the right-moving modes. After 
trivial manipulations and taking into account that 
$T_{p-1}=T_{p}(4\pi^{2}\alpha^{'})^{1/2}$ we can recast $Z_{p}(\beta)$ 
in the more suggestive form
\eqn\corb{
\eqalign{
Z_{p}(\beta)&=-{1\over 8} T_{p-1}^{2}\sum_{i}
\left[{\beta\over 4\pi^{2}\alpha^{'}} 
\sum_{m\in{\bf Z}}\int {d^{25-p}q\over (2\pi)^{25-p}}\right]
\int_{0}^{\infty} ds\, e^{-s\left[q^{2}+{\beta^{2}m^{2}\over 
4\pi^{2}\alpha^{'2}}+
M_{i}^{2}\right]} \cr
&=-{1\over 8} 
T_{p-1}^{2}\sum_{i}\left[{\beta\over 4\pi^{2}\alpha^{'}} 
\sum_{m\in{\bf Z}}\int {d^{25-p}q\over (2\pi)^{25-p}}\right]
\Delta_{i}
}}
where $\Delta_{i}=(q^{2}+M_{i}^{2}+\beta^{2}m^{2}/4\pi^{2}\alpha^{'2})^{-1}$
is the 
propagator for the $i$-th state in the closed bosonic string spectrum. Momentum
conservation in the directions with Neumann boundary conditions sets this 
momentum to zero. On the other hand, momentum
is not conserved in the directions with Dirichlet boundary conditions so we
integrate over the momenta transverse to the D-brane. The interesting 
point is that
the term $\beta^{2}m^{2}/4\pi^{2}\alpha^{'2}$ can be interpreted as 
the discrete momentum squared for a closed string in a circle with length
$\beta_{\rm D}=4\pi^{2}\alpha^{'}/\beta$ \refs\rkap. The compactified 
coordinate has now Dirichlet
boundary conditions, as it is inferred from the non-conservation of the 
discrete momentum in that direction. Therefore the physical picture is that
of a $(p-1)$-dimensional D-brane\foot{If we regard this as a purely euclidean
theory this is the correct picture. However if we think of our euclidean 
theory as the analytic continuation of a Minkowskian string theory the
fact that the time coordinate has Dirichlet boundary conditions means
that the $p$-brane is localized in time, although it can be extended in
space if $p>0$. As a consequence what we have is a $p$-brane whose
world-volume is the $p$-brane itself.} emiting and absorbing closed strings
at inverse temperature $4\pi^{2}\alpha^{'}/\beta$. Actually, we can read off 
from \corb\ the thermal self-energy 
for a static semiclassical Dirichlet $(p-1)$-brane (fig. 1) \fig{Effective
Feynman diagram for 
the self-energy of the bosonic D-brane.}{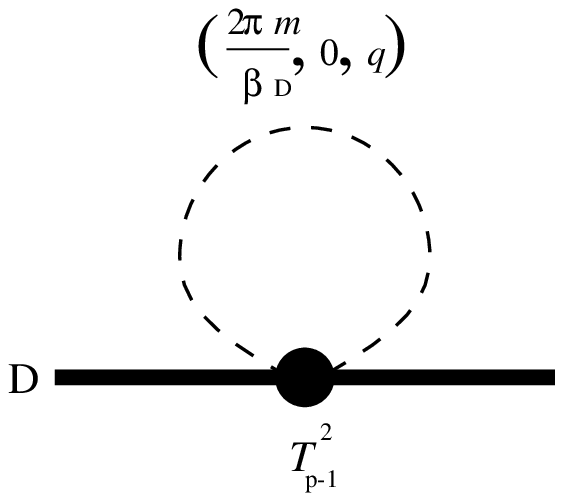}{2truein}
$$
\Pi_{p-1}(\beta)=-T_{p-1}^{2}\sum_{i}\left[{\beta\over 4\pi^{2}\alpha^{'}} 
\sum_{m\in{\bf Z}}\int {d^{25-p}q\over (2\pi)^{25-p}}\right]
\int_{0}^{\infty} ds\, e^{-s\left[q^{2}+{\beta^{2}m^{2}\over 
4\pi^{2}\alpha^{'2}}+
M_{i}^{2}\right]};
$$
therefore we arrive at the following {\it duality} relation between
the Helmholtz free energy of the open bosonic string and the 
D-brane self-energy
\eqn\helmbos{
F_{p}(\beta)={1\over 8\beta}\Pi_{p-1}\left(
{4\pi^{2}\alpha^{'}\over \beta}\right).
}
Notice the appearance of $\beta^{-1}$ in
the prefactor instead of the $\beta^{-2}$ typical of the T-duality of closed 
bosonic strings. As we will see later on
this is linked to the different $\beta\rightarrow 0$ behavior of open 
and closed strings \refs\raw. 

Before leaving the bosonic string let us remark that expression \bos\
admits a D-brane interpretation directly in the open string channel. By 
applying 
a Poisson resumation on the thermal theta function, and using the expressions
of reference
\refs\rrev, one realizes that $\beta F_{p}(\beta)$ is equal to the energy 
of a periodic 
array of $(p-1)$-dimensional D-branes separated by a distance $\beta_{D}
=4\pi^{2}\alpha^{'}/\beta$. Actually, this is nothing more than a trivial 
application of the relation between open strings and D-branes under T-duality,
since by performing a T-duality on the compactified dimension we have now
Dirichlet boundary conditions in that coordinate, increasing the number of
transverse directions to the D-brane and reducing consequently its dimension
from $p$ to $p-1$. This is consistent with our interpretation in the
closed string channel.

The picture found for the bosonic string is not much of a surprise, since
in that case finite temperature amounts to a simple toroidal compactification.
The study of the $SO(N)$, type-I
superstring is by far richer since now   
target fermions in the open string channel are antiperiodic
along the compactified dimension. As we did in the bosonic case, we begin
by transforming \fer\ into the closed string channel. To proceed, however,
it is much more convenient to rewrite $\log Z_{p}(\beta)$ explicitly as
the vacuum energy of a toroidal compactification modded out by the operator
$\alpha=e^{i P_{0}\beta-2\pi i J_{12}}$ \refs\rrohm, where $P_{0}$ is 
the discrete momentum in the euclidean time direction and $J_{12}$ is 
the generator of rotations in the $X^{1}$-$X^{2}$ plane. The result for
the annulus part is\foot{A related study of the annulus partition function 
for the open superstring with $p=0$ can be found in \rgreen.
I thank M.B. Green for attracting my attention to this reference.}
\eqn\ti{\eqalign{
\log Z_{A}(\beta)&=
-{N^{2}\over 8}
\int_{0}^{\infty}{dt\over 2t}(8\pi^{2}\alpha^{'}t)^{-{p\over 2}}
[\eta(it)]^{-12}\left\{[\theta_{3}^{4}(0|it)-\theta_{4}^{4}(0|it)
-\theta_{2}^{4}(0|it)]\,
\theta_{3}\left(0\left|{2i\pi^{2}\alpha^{'}t\over  \beta^{2}}\right.\right)
\right. \cr
&\left.+[\theta_{3}^{4}(0|it)-\theta_{4}^{4}(0|it)
+\theta_{2}^{4}(0|it)]\,
\theta_{4}\left(0\left|{2i\pi^{2}\alpha^{'}t\over  \beta^{2}}
\right.\right)\right\}
}}
The term multiplying the {\it thermal} theta-function $\theta_{3}$ vanishes
identically due to the well-known {\it aequatio identica satis abstrusa}.
In the M\"obius amplitude we find 
\eqn\tmi{\eqalign{
\log Z_{M}(\beta)&=
{N\over 8}
\int_{0}^{\infty}{dt\over 2t}(8\pi^{2}\alpha^{'}t)^{-{p\over 2}}
[\eta(it+1/2)]^{-12} \cr
&\times\left\{[\theta_{3}^{4}(0|it+1/2)-
\theta_{4}^{4}(0|it+1/2)  -\theta_{2}^{4}(0|it+1/2)]\,
\theta_{3}\left(0\left|{2i\pi^{2}\alpha^{'}t\over  \beta^{2}}
\right.\right) \right.  \cr
&\left.+[\theta_{3}^{4}(0|it+1/2)-\theta_{4}^{4}(0|it+1/2)
+\theta_{2}^{4}(0|it+1/2)]\,
\theta_{4}\left(0\left|{2i\pi^{2}\alpha^{'}t\over  \beta^{2}}\right.
\right)\right\}.
}}

Let us study first the orientable part of the amplitude. As we did in 
the bosonic case, we rewrite \ti\ in the closed string channel by 
changing variables $s=t^{-1}$. After this we can identify the different
sectors in the new channel; $\theta_{3}^{4}(0|is)-\theta^{4}_{4}(0|is)$ 
corresponds to closed string states in the NS-NS sector with 
$G=G_{L}=G_{R}=+1$,
whereas $\theta_{2}^{4}(0|is)$ represents the contribution of R-R states
with $G=G_{L}=\pm G_{R}=+1$; $G$ is the $G$-parity of the closed string 
state and $\pm G_{L}$ for the R-R states applies respectively when $p$ is 
even or odd\foot{This is because
type-IIB closed strings couple to odd-dimensional D-branes while those of the
type-IIA do to even-dimensional ones. We will see in a moment that in the 
closed string channel we deal with $(p-1)$-dimensional D-branes.}.
In the same manner, $\theta_{3}^{4}
(0|is)+\theta^{4}_{4}(0|is)$ and $\theta_{2}^{4}(0|is)$ in the second
part of the expression correspond respectively to NS-NS and R-R with 
$G=-1$. These latter are the states absent in the
supersymmetric closed string theory and their contributions to the partition
function disappear in the zero temperature limit. 
By rescaling the integration variable and identifying the mass formulae
in the different sectors we arrive at
\eqn\annulus{\eqalign{
&\log Z_{A}(\beta)=-{1\over 16}{N^{2}\over 2}
T_{p-1}^{2}\left[{\beta\over 2\pi^{2}\alpha^{'}}\sum_{m\in {\bf Z}}
\int{d^{9-p}q\over (2\pi)^{9-p}}\right]\int_{0}^{\infty}ds\,\left[
\sum_{\rm (NS,+)}e^{-s\left[q^{2}+{m^{2}\beta^{2}\over \pi^{2}\alpha^{'2}}+
M^{2}_{i}\right]}\right.\cr
&\left.-\sum_{\rm (R,+)}e^{-s\left[q^{2}+{m^{2}\beta^{2}\over 
\pi^{2}\alpha^{'2}}+M^{2}_{i}\right]}+
\sum_{\rm (NS,-)}e^{-s\left[q^{2}+{(m+1/2)^{2}\beta^{2}\over 
\pi^{2}\alpha^{'2}}+M^{2}_{i}\right]}
-\sum_{\rm (R,-)}e^{-s\left[q^{2}+{(m+1/2)^{2}\beta^{2}\over 
\pi^{2}\alpha^{'2}}+M^{2}_{i}\right]}\right].}}
The sums inside the integral correspond to the four types of closed
string states, NS-NS and R-R with positive and negative $G$-parities. 
As it is the case in at zero temperature the contributions
from NS-NS and R-R have opposite signs but now they cancel each other only
in the $G=+1$ sector. In fact,  a very interesting
picture for the orientable part of the ``hot'' D-brane emerges from \annulus.
Because of
the form of the discrete contributions to the propagator the closed
strings can be thought to be living in a circle with length 
$\beta_{\rm D}=2\pi^{2}\alpha^{'}/\beta$. As in the bosonic string, 
now $X^{0}$ has Dirichlet boundary conditions, momentum is not conserved
in that direction and we deal with a 
$(p-1)$-dimensional D-brane. Moreover, odd $G$-parity states have
half-integer momentum numbers so they are treated as antiperiodic in 
the compactified dimension, or in other words, they behave as fermions
from the thermal point of view \refs\rkap. Since the $G$ parity operator
is essentially the world-sheet fermion number the final conclusion is that
in going from the open string to the D-brane picture we are, in a sense, 
trading the r\^oles of target and world-sheet statistics in the finite 
temperature setup. As in the case of the bosonic string one read from 
\annulus\ the corresponding Feynman diagram in the effective D-brane theory 
and find that it can also be interpreted as the self-energy of the system
of $N$ D-branes where odd $G$-parity 
bosonic states are twisted (they are treated as fermions). This is equivalent
to mod out the tree level closed string spectrum by the operator 
$\tilde{\alpha}=e^{i \tilde{P}_{0} \beta_{D}}G$, with $\tilde{P}_{0}$
the momentum in the dual circle with length $\beta_{D}$.  

One remarkable thing about this picture is the emergence of the 
dual temperature under $\beta$-duality
as the ``temperature'' of the Dirichlet $(p-1)$-brane,
differing from the value implied by T-duality by a factor two.
Then the annulus canonical free 
energy $F_{A}(\beta)$ can be written in terms of the self-energy of the 
system of coincident D-branes at temperature ${2\pi^{2}\alpha^{'}/\beta}$
$$
F_{A}(\beta)= {1\over 16\beta}
\Pi_{p-1}\left(2\pi^{2}\alpha^{'}\over \beta\right).
$$

Let us move to the M\"obius amplitude. Now the closed string channel will
describe a closed string emmited by the D-brane which ends in a crosscap
located in the vicinity of the orientifold plane. Therefore what we expect
to obtain is a description of the M\"obius free energy in terms of the
self-energy of the system D-brane-orientifold (fig. 2). \fig{Feynman diagram 
for the self-energy of the system D-brane-orientifold.}{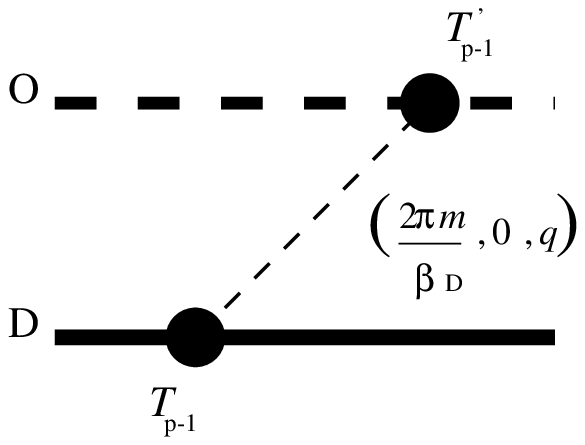}{2truein}
In order to go from the 
open to the closed string channel it is useful to re-express \tmi\
in terms of modular functions depending on $2t$. The closed string
channel is gotten then 
by the replacement $2t \rightarrow (2s)^{-1}$. Skipping the
details we get
$$\eqalign{
\beta F_{M}(\beta)&=
{N\over 8} 2^{p-7\over 2}(4\pi^{2}\alpha^{'})^{-{p+1\over 2}}\beta
\int_{0}^{\infty} ds \, s^{-{9-p\over 2}}
[\eta(is+1/2)]^{-12} \cr
&\times\left\{[\theta_{3}^{4}(0|is+1/2)-
\theta_{4}^{4}(0|is+1/2)  -\theta_{2}^{4}(0|is+1/2)]\,
\theta_{3}\left(0\left|{2i\beta^{2}s\over \pi^{2}\alpha^{'}}
\right.\right) \right.  \cr
&\left.+[\theta_{3}^{4}(0|is+1/2)-\theta_{4}^{4}(0|is+1/2)
+\theta_{2}^{4}(0|is+1/2)]\,
\theta_{2}\left(0\left|{2i\beta^{2}s\over \pi^{2}\alpha^{'} }\right.\right)
\right\}.}
$$ 
The identification of R-R and NS-NS states now
is a little more involved. From \refs\rcaipol\ we learn that $\theta_{3}^{4}-
\theta_{4}^{4}$ corresponds to R-R, while $\theta_{2}^{4}$ is
the contribution of NS-NS states, both with $G=+1$. Taking this into account 
and using some well-known properties of the Jacobi theta-functions 
\refs\rwguo\ one arrives, after a redefinition of the proper time, at
\eqn\moebius{\eqalign{
&\log Z_{M}(\beta)=-{1\over 2}{N\over 2}
T_{p-1}T_{p-1}^{'}\left[{\beta\over 2\pi^{2}\alpha^{'}}\sum_{m\in {\bf Z}}
\int{d^{9-p}q\over (2\pi)^{9-p}}\right] \cr
&\times\int_{0}^{\infty}ds\,\left[
\sum_{\rm (NS,+)}(-1)^{m+N_{i}}
e^{-s\left[q^{2}+{m^{2}\beta^{2}\over \pi^{2}\alpha^{'2}}+
M^{2}_{i}\right]}
-\sum_{\rm (R,+)}(-1)^{N_{i}}e^{-s\left[q^{2}+{m^{2}\beta^{2}\over 
\pi^{2}\alpha^{'2}}+M^{2}_{i}\right]}\right]}}
where $N_{i}$ is the total right-moving oscillator number
and we have identified the tensions of the D-brane and the orbifold plane,
$T_{p}$ and $T_{p}^{'}$. 
First of all we find that, in contrast with the annulus case,
in the M\"obius partition function we have
contributions only from the even $G$-parity bosonic closed string states. 
In addition, from inspection of the exponents, we see that their
discrete momenta correspond to those of closed strings living in a
circle with length $\beta_{D}=2\pi^{2}\alpha^{'}/\beta$, again the 
$\beta$-duality value that we also got for the annulus. 
It is interesting to notice that at finite temperature the cancellation
between the R-R and NS-NS states only takes place for those states
with even momentum number, whereas for odd $m$  both contributions
add up to give a net non-vanishing result.
The general structure of \moebius\ is 
$$
\log Z_{M}(\beta)=-{1\over 2}{N\over 2}
T_{p-1}T_{p-1}^{'}\sum_{i}
\left[{\beta\over 2\pi^{2}\alpha^{'}}\sum_{m\in {\bf Z}}
\int{d^{9-p}q\over (2\pi)^{9-p}}\right] \Omega_{i}\Delta_{i}
$$
where again $\Delta_{i}$ is the propagator for the $i$-th state of the 
type-II superstring. $\Omega_{i}$ is a ``twist'' operator in the closed
string channel that turns out to be
$$
\Omega_{i}={1+G_{i}\over 2} e^{i\tilde{P}_{0}\beta_{D}a} (-1)^{N_{i}}
$$
with $a=1/2,0$ for NS-NS and R-R states respectively; 
the projector selects the states
with $G=+1$, the only ones coupling to the orientifold plane. 
This provides us with the ``finite temperature'' prescription
for constructing the self-energy of the system  $(p-1)$-brane-orientifold,
with the result
$$
F_{p}(\beta)={1\over 2\beta}\Pi_{p-1}^{\rm D-O}
\left({2\pi^{2}\alpha^{'}\over \beta}
\right).
$$

For the unoriented superstring an open string channel interpretation of \fer\
along the lines of the one offered for the bosonic string can also be made.
In this case what we find is an array of $(p-1)$-dimensional D-branes
located at $Y_{D}=2\pi^{2}\alpha^{'}n/\beta$, while the orientifold planes 
are at $Y_{O}=\pi^{2}\alpha^{'}n/\beta$, with $n\in{\bf Z}$. The annulus 
amplitude describes
open strings stretched between two D-branes with a modding impossing that
bosonic modes wrap a even number of times whereas the ``winding'' number
for the fermionic ones is odd. For the M\"obius amplitude the scenario 
is similar where now the endpoints of the string are located at a give 
D-brane and its mirror image with respect to the orientifold.

\newsec{The high temperature limit}

Once we have arrived at a D-brane description of open strings at finite
temperature there are a couple of things that can be studied from the new point
of view. The first one is the physical meaning of the Hagedorn temperature.
As we already indicated in the introduction, this is a problem that one
does not hope to solve in the D-brane scenario; however it is of some 
interest to see how the divergence arise in the dual picture. 

In both the bosonic and the fermionic string the one-loop partition function in
the closed string channel is written as the trace over closed string
propagators at zero momentum in the spatial directions with Neumann boundary
conditions. From \corb, \annulus\ and \moebius\ we verify that divergences
will arise whenever we have a tachyonic state for small $\beta$. Thinking in 
terms of D-branes, the Hagedorn divergence arises for small values of the 
dual ``temperature'' triggered by the self-interaction of the D-brane via 
a tachyon. This divergence is removed at high dual 
temperatures because then the effective mass squared of the closed string 
tachyons becomes positive. This is much more reminiscent of a 
field-theoretical scenario, in which tachyonic ground states are removed at 
high temperature by symmetry restoration \refs\rdoljack. In the annulus
partition function there are tachyons in the bosonic string and in the
$G=-1$ NS-NS sector of the type-I superstring. In the M\"obius partition 
function, on the contrary, there are no tachyons since it 
only involves the $G=+1$ sector of the type-II superstring, so its 
contribution is regular for all values of the temperature. In the 
open string channel the Hagedorn transition is marked by a winding open
string becoming tachyonic at high temperatures \refs\rgreen.

Another interesting issue is the formal $T\rightarrow \infty$ behavior.
For the closed ten-dimensional superstrings the canonical free energy 
goes like $\beta^{-2}\Lambda$ where $\Lambda$ is a constant that in the 
case of the heterotic string coincides with the vacuum energy at $T=0$ 
\refs\raw. 
This is the asymptotic behavior to be expected from
a two-dimensional field theory and the conclusion is that heterotic
strings have very few propagating states at high temperature.
For the open bosonic string in the presence of a $p$-brane we can obtain the 
$\beta\rightarrow \infty$ using general arguments. In section 2 we 
constructed $F_{p}(\beta)$ by starting with the $D$-dimensional open string 
with all spatial dimensions compactified in circles
and taking the limit $R_{i}\rightarrow 0$ in those transverse to the 
D-brane world-volume. If we now also take the radius of the euclidean time
to zero we will end up with the vacuum energy of open strings in the background
of a $(p-1)$-dimensional D-brane at zero temperature. This is exactly the
result found by computing the limit explicitly\foot{I am indebted to
M.A.R. Osorio for sharing with me his unpublished results on the high 
temperature limit of open superstrings.} after regularizing the
integral with an ultraviolet cuttoff $\epsilon$ (cf. \refs\raw)
\eqn\htl{
F_{p}(\beta)\rightarrow -{1\over \beta}\int_{\epsilon}^{\infty}
{dt\over 2t}(8\pi^{2}\alpha^{'}t)^{-{p\over 2}}[\eta(it)]^{-24}=
{\Lambda_{\epsilon}\over \beta}
}
From the open string analog model point of view, $\Lambda_{\epsilon}$
is nothing but the ultraviolet regularized vacuum energy of the whole
collection of quantum fields in the open string attached to a
$(p-1)$-brane. Actually, it is easy to translate \htl\ to the 
closed string channel by going to \corb\ and replacing the discrete sum
by an integral over a continuun momentum, so now we have a single 
$[25-(p-1)]$-dimensional integral. There, as expected, 
the divergence when $\epsilon$ goes to zero is due to the closed string 
tachyon propagating in a long tube; in going to the closed string channel
we are trading the ultraviolet cuttoff $\epsilon$ by an infrared one, 
$1/\epsilon$.

The situation with the superstring is complicated by 
funny boundary conditions. Since space-time fermions are antiperiodic
in the timelike direction, in the limit $\beta\rightarrow 0$ all fermionic
degrees of freedom dissappear and we are left only with space-time bosons.
This is true for both the annulus and the M\"obius amplitude. In the D-brane
picture the ``temperature'' goes to infinity and therefore the momentum in
the time direction becomes continuum. So in equations \annulus\ and \moebius\
the sum over the integer is replaced by an integral over a momentum $q_{0}$
and consequently, as with the bosonic string, we end up with an integral 
over a $[9-(p-1)]$-dimensional momentum. In the case of the annulus the 
integrand is the trace over the four bosonic sectors of 
$\exp{[-s(q^{2}+M_{i}^{2})]}$. In the
M\"oebius amplitude,  the alternate sign on the NS-NS sector makes this
part to dissapear in that limit so we are left only with the trace of the same
exponential to the positive $G$-parity R-R states\foot{This is a consequence
of the fact NS-NS states in the M\"obius strip correspond in the open string 
channel to R target fermions.}. The corresponding $\beta\rightarrow
0$ limits are
$$
\eqalign{F_{A}(\beta)&\rightarrow -{1\over \beta}{N^{2}\over 4}
\int_{\epsilon}^{\infty}
{dt\over 2t} (8\pi^{2}\alpha^{'}t)^{-{p\over 2}}{\theta_{3}^{4}(0|it)-
\theta_{4}^{4}(0|it)\over \eta^{12}(it)} \cr
F_{M}(\beta)&\rightarrow {1\over \beta}{N\over 4}
\int_{\epsilon}^{\infty}
{dt\over 2t} (8\pi^{2}\alpha^{'}t)^{-{p\over 2}}{\theta_{3}^{4}(0|it+1/2)-
\theta_{4}^{4}(0|it+1/2)\over 	\eta^{12}(it+1/2)}
}$$
It is easy to realize that the coefficients of $\beta^{-1}$, once 
transformed to the closed string channel, can be intepreted as the
modded ``zero temperature'' self-energy of the D-brane. The 
regularization of the 
integrals is needed in the case of the annulus amplitude, since we have
a divergence associated with the closed string NS-NS tachyon. 
In the M\"obius case,
however, there is no tachyon in the closed string channel and therefore
any infrared divergence can only be due to massless modes; indeed, 
the integral will be infrared finite in the closed string channel
($t\rightarrow 0$) whenever $p<8$. Ultraviolet convergence, on the other
hand, is guaranteed by the tower of massive states. 

This high temperature behavior might seem somewhat 
extrange. After all, the absence of the modular group $SL(2,{\bf Z})$
in the open string at one-loop 
is usually taken as an indication that open string theory is more or less
equivalent to a collection of quantum fields. This being so, it seems
natural that in the limit $\beta\rightarrow 0$ we should have
$F_{p}(\beta)\sim \beta^{-(p+1)}$. There is however a very important 
point to remember: the asymptotic
behavior $\beta^{-D}$ for the one-loop free energy of quantum fields in a 
$D$-dimensional space-time is gotten provided that there is no ultraviolet 
divergences in the free energy other that the ones already present in the 
zero-temperature theory. Although this condition is fulfilled individually 
for each field in the string spectrum this is not the case for the collection 
as a whole. In the string partition function we have the Hagedorn divergence
triggered by the exponential growth of the number of states per mass level;
this infinity appears in the ultraviolet region in the proper time
representation of $F_{p}(\beta)$. It is necessary then to regularize the 
ultraviolet behavior of the integral in order to extract some sensible
result when the temperature goes to infinity and therefore we have to 
pick up a regularization length scale $\Lambda$. The most natural choice
is to relate this scale to the string scale by defining $\Lambda^{2}=
\epsilon\,\alpha^{'}$. Then in the limit $\beta\rightarrow 0$ we have
$$
F_{p}(\beta)\sim {1\over \beta\Lambda^{p}} \sum_{i} 
f(\Lambda M_{i})+O(e^{-{\Lambda^{2}\over \beta^{2}}})
$$
where the sum is over an appropriate subset of the string spectrum.
This is exactly what we have been doing in this section.
Had we redefined the proper time $t\rightarrow \beta^{2}t/\alpha^{'}$
in equations \bos\ and \fer\ and placed an ultraviolet cuttoff $\epsilon$
(the procedure which gives us the $\beta^{-D}$ behavior for a single 
quantum field), the final result would be $\beta^{-(p+1)}g(\beta^{2}/\alpha^{'}
)$ where now, however, $g(x)$ has no well-defined expansion around $x=0$.

\newsec{Conclusions}

In this letter we have studied finite temperature dualities for open strings
in the presence of  $p$-dimensional D-branes. By writting the one-loop
canonical free energy in the closed string channel we re-expressed 
$F_{p}(\beta)$ for the bosonic string without Chan-Paton factors
in terms of the self-energy of a semiclassical $(p-1)$-dimensional 
Dirichlet brane at inverse temperature
$4\pi^{2}\alpha^{'}/\beta$. In the case of the type-I, $SO(N)$ superstring the 
annulus free energy is given by the self-energy of a system of $(p-1)$-branes 
at $\beta_{D}=2\pi^{2}\alpha^{'}/\beta$, where the r\^ole of the space-time 
statistic is played by the $G$-parity of the closed string state. For the 
M\"obius part the self-energy of the system D-brane-orientifold receives 
contributions only from the even $G$-parity closed string states
although they are affected by phases coming from the ``twist'' 
operator. 

Remarkably, in both the annulus and the M\"obius type-I amplitudes
the dual temperature felt by the closed strings coupled to the D-brane
is determined by $\beta$-duality, $2\pi^{2}\alpha^{'}/\beta$. 
It is very importat to stress that, either working with $\alpha^{'}_{\rm open}$
or $\alpha^{'}_{\rm closed}$, what caracterizes $\beta$-duality is the
factor one half relative to the 
corresponding dual length under T-duality. Although $\beta$-duality 
is well-known in the context of heterotic strings at finite temperature,
in fact it has also appeared in non-heterotic contexts,
for example in the formal temperature duality of type-II superstrings
\refs\raw\ros. Its emergence also in the open superstring scenario strongly
suggest that $\beta$-duality has to be viewed as a correction of T-duality
in toroidal compactifications of any fermionic string with antiperiodic target 
fermions. It would be interesting to clarify the physical origin of
such correction.

We have also studied the emergence of the Hagedorn singularity in the dual
picture and verified that it is triggered by the emision of tachyonic closed 
string states by the D-brane. The absence of this singularity in the M\"obius
amplitude steems from the absence of tachyonic interchanges between the
D-brane and the orientifold plane. Studying the formal limit $\beta\rightarrow
0$ we find that the one-loop canonical free energy diverges as 
$\Lambda_{\epsilon}\beta^{-1}$ in contrast with the closed string behavior 
as $\beta^{-2}$ \refs\raw. The regularized constant $\Lambda_{\epsilon}$ is 
interpreted in the closed string channel as the ``zero temperature'' 
self-energy of the D-brane coupling to NS-NS and R-R closed string states
with $G=\pm 1$
(or of the D-brane-orientifold system interchanging only R-R states with $G=1$ 
in the case of the M\"obius amplitude). In the 
open string channel this same constant corresponds to the vacuum energy of
open strings in the presence of a 
$(p-1)$-dimensional D-brane with target fermions eliminated.
This is just a stringy version of dimensional reduction at high temperature.

\newsec{Acknowledgements}

I am grateful to J.L.F. Barb\'on and M.A.R. Osorio for enlightening 
discussions and comments and to M.B. Green for useful remarks on a
first version of this paper. This work has been supported by a 
MEC (Spain) postdoctoral fellowship. This work is dedicated to Mar\'{\i}a Ruiz
V\'azquez in her 80th birthday.

\listrefs
\bye